# The vacuum thermal treatment effect on the optical absorption spectra of the TiO$_2$ coated by Ni-B nano-clusters photocatalyst powders

**M.M. Nadareishvili, K.A. Kvavadze, G.I. Mamniashvili*, T.N. Khoperia, T.I. Zedgenidze**

E. Andronikashvili Institute of Physics, Tamarashvili 6, 0177 Tbilisi, Georgia

*Corresponding author.
E-mail address: g.mamniashvili@aiphysics.ge

## ABSTRACT

The thermal vacuum treatment effect on the optical absorption spectra of the TiO$_2$ nanopowders, both pure and coated by the Ni-B clusters with the original electroless method was investigated. It was observed that the thermal treatment of pure TiO$_2$ nanopowders does not change their optical absorption spectrum while after the coating of these powder particles by the Ni-B clusters the thermal treatment results in the increase of the optical light absorption in the visual region of spectrum. This points to the possibility of the significant improvement of the photocatalist efficiency of TiO$_2$ nanopowders coated by the Ni-B clusters using the thermal treatment.

## 1. Introduction

It is well known that gas and fuel supply on the earth at conditions of current consuming will be soon exhausted. The current fuel consumption intensity is defined by the industrialization of mankind and is not practically possible to slow it down. For this reason the only way out from this dependence is the finding out alternative energy sources: one of the most perspective direction on this way is the water dissociation on hydrogen and oxygen using solar energy and utilization of produced hydrogen as fuel, final product of which under burning is again water.

The nowadays actual problem is the increase of efficiency of photocatalytic reaction – the reaction of water dissociation on hydrogen and oxygen by the way of solar radiation energy application using catalysts [1-9]. It is believed currently that the most perspective substance for this aim is the titanium dioxide $TiO_2$. Photocatalysis over a semiconductor oxide such as $TiO_2$ is initiated by the absorption of a photon with energy equal to or greater than the band gap of the semiconductor (3.2 eV for $TiO_2$) producing electron – hole ($e^-/h^+$) pairs [9].

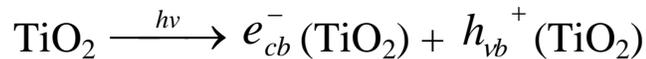

$$TiO_2 \xrightarrow{\ hv\ } e^-_{cb}(TiO_2) + h^+_{vb}(TiO_2)$$

Consequently, following irradiation, the $TiO_2$ particle can act as either an electron donor or acceptor for molecules in the surrounding media (Fig. 1). However, the photoinduced charge separation in bare $TiO_2$ particles has a very short lifetime because a charge recombination. Therefore, it is important to prevent hole-electron recombination before a designated chemical reaction occurs on the $TiO_2$ surface.

Having recognized that charge separation is a major problem, numerous techniques were invented to minimize this effect. One such technique is to scavenge photogenerated charges with strongly adsorbed species. Increased charge separation distance can be obtained adding a metal and a metal oxide clusters to the surface of the semiconductor particle. Such a particle is shown below (Fig. 2).

The electrons produced upon band-gap excitation are injected into the metal particles, and positively charged holes are injected into the metal oxide.

Alternatively, a sacrificial species can be used to remove either holes or electrons, allowing them to react. As it was pointed out, a band-gap width for $TiO_2$ is 3.2 eV, therefore only ultra-violet rays participate in the catalitic reaction which share in the solar radiation spectrum is only about 4 %. Necessary energy for water dissociation is 1.23 eV. The energy of corresponding photons is situated in the visible part of spectrum which energy on the order of value greater as compared with the ultra-violet region. Therefore to improve of the catalytic reaction efficiency it is important to increase the contribution of visible part of light spectrum in the catalytic process.

## 2. Results and discussions

In the Andronikashvili Institute of Physics it was developed the original method of coating nano-particles by clasters of different sizes from different materials (as example, Ni-B) [10]. These nanostructures were fabricated using electroless deposition of metals and alloys. The peculiarity of method is in the maintaining of low-temperature during the coating reaction $(58 - 60^{o}C)$.

We made some previous investigations in this direction, namely, it was found out definite methods which gave possibility to change optical properties of above mentioned Ni-B/TiO$_2$ powders in the prescribed direction. For this aim it was studied the optical absorption spectra of the photocatalyst powder distillate suspense prepared by the above mentioned methods. Two crystallographic modifications of TiO$_2$ powder (anatase and rutile) were used. The light absorption of distilled water over the entire working spectral range was preliminary studied. It appeared to be rather low. However, in order to exclude the distortion of absorption spectra of the powders under study as a result of the distillate effect, we recorded the absorption spectra of the powders dissolved in the distillate in reference to pure distillate instead of the air. For this purpose the cell with the solution of the powder under study was placed in one compartment of the 4-compartment spectrophotometer, and the identical cell with pure distillate – in other compartment. To check the validity of the used procedure, we prepared several similar reference aqueous solutions of the same powder. Then their absorption curves were taken. The obtained spectra coincide very closely with each other with the maximal difference of $\pm$ 5%. In Fig. 3 it is shown two typical examples of above mentioned curves.

In this paper it is investigated the heat treatment effect on the optical absorption spectra of the TiO$_2$ nano-powders coated by Ni-B nano-clusters by the above mentioned original method and the improvement of their photocatalitic properties, by increasing of visible light share in the photocatalitic process and, correspondingly, for improvement of photocatalysis reaction efficiency by this method.

Firstly the heat treatment effect on the optical absorption spectra of pure TiO$_2$ (both anatase and rutile) powders was studied. The size of powders particles was ~ 500 nm. The absorption peaks of these powders are mainly situated in the ultra-violet region and at thermal vacuum treatment the absorption spectra of these powders practically remain unchanged. As per the coating of these powders by Ni–B clusters, after this procedure the absorption spectrum of rutile was not practically changed again but the absorption of anatase in the visible part of spectrum was increased.

In Fig. 4 absorption spectra before a thermal treatment and after it for $TiO_2$ (anatase) coated by Ni–B clusters are shown. The curve 2 corresponds to optical absorption spectrum thermally untreated Ni - B/$TiO_2$ (anatase) powder suspension, and the curve 1 – to thermally treated Ni–B/ $TiO_2$ (anatase) powder. As it is seen from the figure that as a result of thermal treatment one could observe the significant change of optical absorption spectrum of Ni–B/ $TiO_2$ powder suspension, namely, in the optical absorption spectrum a new peak appeared on wave length ~ 350 nm and the absorption of solar energy significantly increases in the range of 300 – 400 nm. This fact is particularly important for increasing of the photocatalist efficiency in the visual range of solar radiation.

So, it follows from our experiments that the thermal treatment of $TiO_2$ nanopowders does not change their electron excitation spectrum. But after the coating of $TiO_2$ by the Ni–B clusters the thermal treatment results in the change of the electron excitation spectrum of these particles followed by the increase of absorption in the visual part of spectrum.

In the summary, with the help of the thermal treatment and maintaining corresponding modes of thermal treatment, it is possible to improve significantly the photocatalist efficiency of the $TiO_2$ powders coated by the Ni – B nanoclasters.

The work is supported by the joint GNSF/STCU Grant No. 4677.

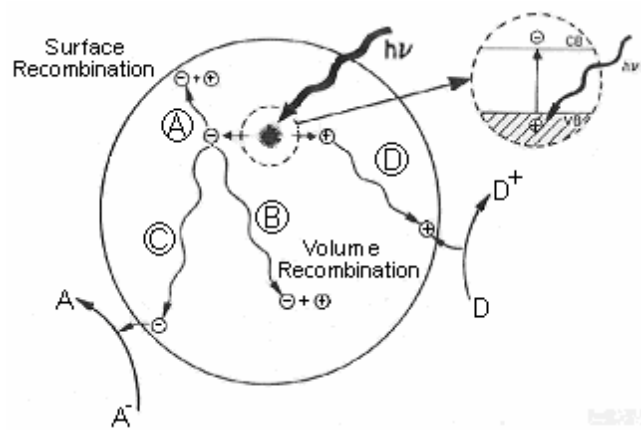

Fig. 1.

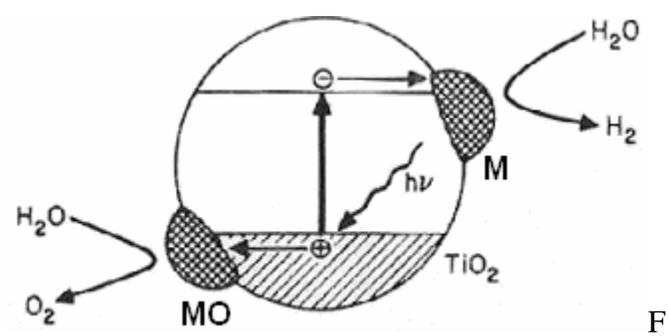

Fig. 2.

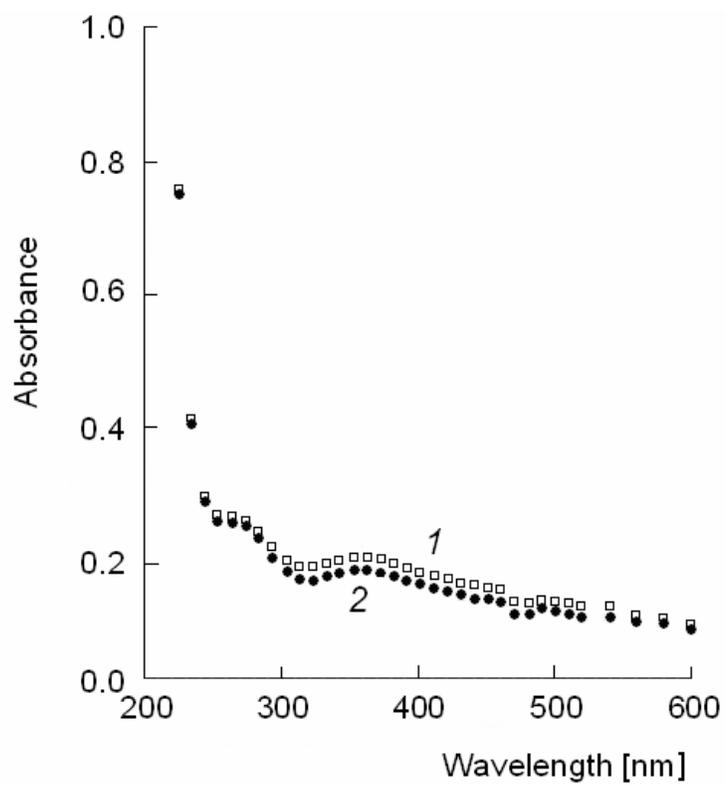

Fig. 3.

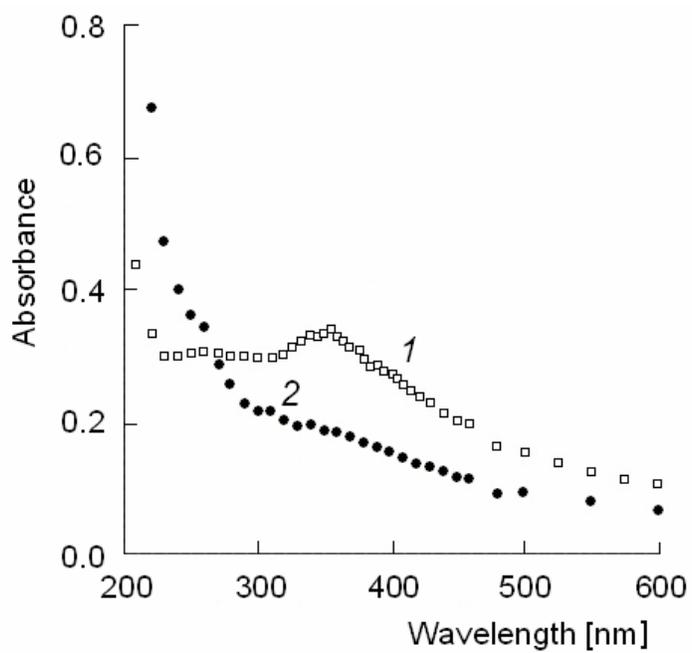

Fig. 4.

FIGURE CAPTIONS

Fig. 1. Irradiation of the $TiO_2$ particles.

Fig. 2. Charge separation by the metal (M) and metal oxide (MO) clasters on the surface of the semiconductor particle.

Fig. 3. Absorption curves of two different reference solutions

Fig. 4. Absorption spectra for $TiO_2$ (anatase) photocatalists coated by Ni – B clasters after the thermal treatment (curve 1) and before it (curve 2).